\documentclass[%
 reprint,
 amsmath,amssymb,
 aps,
]{revtex4-2}

\usepackage{graphicx}% Include figure files
\usepackage{subcaption}
\usepackage{dcolumn}% Align table columns on decimal point
\usepackage{bm}% bold math
\newcommand{\Lir}{\hat{L}^{ir}}

\newcommand{\LFP}[1]{\hat{L}^{FP}_{(#1)}}

\newcommand{\Cvv}{\boldsymbol{C}^{vv}}
\newcommand{\Cff}{\boldsymbol{C}^{ff}}
\newcommand{\Ctilde}{\tilde{\boldsymbol{C}}^{vv}}

\newcommand{\Top}{\hat{\mathcal{T}}}

\newcommand{\Rop}[1]{\hat{\mathcal{R}}_{#1}}

\newcommand{\Liou}[1]{\hat{L}_{(\boldsymbol{X},#1)}}

\newcommand{\atB}{{}_{(B)}}
\newcommand{\atmB}{{}_{(-B)}}

\begin{document}

\preprint{APS/123-QED}

% \title{From Langevin to Brownian dynamics with the Lorentz force: \\ The interplay  between velocity and force autocorrelation}% Force line breaks with \\

\title{Velocity and force autocorrelations in Brownian dynamics with a Lorentz force}

\author{Filippo Faedi}
 \affiliation{University of Augsburg, Institute of Physics, D-86159 Augsburg, Germany.}%Lines break automatically or can be forced with \\
\author{Abhinav Sharma}%
 \email{abhinav.sharma@uni-a.de}
\affiliation{University of Augsburg, Institute of Physics, D-86159 Augsburg, Germany.}%
\affiliation{Leibniz-Institute for Polymer Research, Institute Theory of Polymers, D-01069 Dresden, Germany}

\begin{abstract}
We derive a general relation between the velocity and force autocorrelation tensors (VACT 
and FACT) for a Brownian particle subject to an external magnetic field. Using 
time-symmetry arguments, we show that, for the full Langevin dynamics, the VACT depends 
only on the FACT, independently of the details of the interaction potential. Under the 
hypothesis of timescale separation between thermalization and interaction-driven motion, 
this relation simplifies considerably in the overdamped (Brownian) limit. A central 
feature of the overdamped result is that, unlike in the field-free case, the part of the 
VACT that controls the self-diffusion of the particle couples to the antisymmetric part of 
the FACT, with a coupling strength set by the ratio of the cyclotron frequency to the 
thermalization rate. We validate and illustrate the formalism on an exactly solvable model: 
a dimer of charged particles bound by a harmonic potential. Depending on the relative sign 
of the particle charges, the magnetic field is found to produce either a transient 
suppression of mobility and diffusion that is fully recovered at long times, or a 
persistent oscillatory force autocorrelation, regions of negative mobility, and a 
 long-time suppression of self-diffusion.
\end{abstract}

%\keywords{Suggested keywords}%Use showkeys class option if keyword
                              %display desired
\maketitle
\onecolumngrid
%\tableofcontents
\section{Introduction}

The dynamics of Brownian particles in a viscous solvent is well described by the Langevin 
equation, which captures the interplay between deterministic forces, hydrodynamic friction, 
and thermal noise. When inertial effects are negligible compared to viscous damping, one 
works in the overdamped (Brownian dynamics) limit, in which the velocity degrees of freedom 
are formally eliminated. Nevertheless, it remains possible to reconstruct the velocity 
autocorrelation tensor (VACT) from purely configurational quantities. Specifically, the VACT 
can be expressed in terms of the force autocorrelation tensor (FACT) via
\begin{equation}\label{VACT_BROWNIAN_2}
    \gamma^2\left<\boldsymbol{v}(t)\otimes\boldsymbol{v}(0)\right> = 
    \gamma k_B T \,\boldsymbol{I}\, \delta^+(t) 
    - \left<\boldsymbol{f}(t) \otimes\boldsymbol{f}(0)\right>,
\end{equation}
where $\gamma$ is the friction coefficient, $k_BT$ the thermal energy, and $\boldsymbol{f}$ 
the conservative interparticle force. This relation has proven to be a powerful tool in both 
analytical studies and Brownian dynamics simulations of interacting colloidal suspensions, 
as it encodes how interactions shape particle mobility and, through the Green--Kubo relation, 
determines the self-diffusion coefficient~\cite{hanna1981velocity, sharma2016communication,Franosch2019, Franosch2026}.

The standard derivation of Eq.~\eqref{VACT_BROWNIAN_2} relies on the time-reversal symmetry 
of the equilibrium dynamics. This symmetry is broken when the particles carry an electric 
charge and are subject to an external magnetic field, because the resulting Lorentz force is 
velocity-dependent and changes sign under time reversal~\cite{Akesson1985}. The breakdown of 
time-reversal symmetry has an immediate consequence for the derivation: the standard 
arguments no longer apply, and the relation between the VACT and the FACT must be 
reconsidered from the outset.

Beyond the formal complication, a fundamental physical question arises: does the Lorentz 
force have any measurable influence on the dynamics in configurational space, or is its 
effect washed out by thermal fluctuations? Since the Lorentz force depends explicitly on 
the particle velocity, and the velocity is strongly randomized by collisions with the 
solvent, it is not a priori clear whether any systematic effect survives. Chun et 
al.~\cite{Non_white_noise} showed that the answer depends sensitively on the assumed 
timescale separation at the level of the Langevin equation. The relevant dimensionless 
parameter is $\kappa = \omega_c \tau_\mathrm{th}$, where $\omega_c$ is the cyclotron 
frequency and $\tau_\mathrm{th}$ is the thermalization time of the particle. In the limit 
of fast thermalization, $\tau_\mathrm{th} \to 0$, the Lorentz force does leave a trace in 
the configurational dynamics, provided the product $\kappa$ remains finite -- a condition 
realizable with highly charged colloidal particles in strong magnetic fields. This regime 
is directly relevant to experiments on dusty (complex) plasmas, for which dedicated 
experimental apparatus have recently been constructed~\cite{Thomas_2012, thomas2015magnetized, ludwig2012wake, melzer2021physics, jaiswal2017effect, choudhary2024magnetized, Rosenberg_2014,Tarasov}.

Turning to interacting many-body systems, prior work on chiral and magnetically driven 
fluids has revealed that the Lorentz force can have striking consequences even in the 
overdamped limit. Kalz et al.~\cite{Osc_Forc_Autcorr} demonstrated, for a system of overdamped 
interacting particles with hard-sphere interactions in an external magnetic field, that the 
FACT exhibits oscillatory, nonmonotonic decay in time -- a behavior that is highly 
unexpected for overdamped dynamics, where one would normally anticipate a monotonic 
relaxation. Through the VACT--FACT relation, these oscillations feed into the 
Green--Kubo integral and produce an enhancement of the self-diffusion coefficient relative 
to the field-free case~\cite{ shinde2022strongly, Coll_Enh_Self_Diff, langer2024dance}. On the underdamped 
side, the tensor structure of the VACT and FACT in the presence of a magnetic field, and 
the time-symmetry constraints they must satisfy, were analyzed by Coretti et 
al.~\cite{Coretti1, Coretti2} for Hamiltonian dynamics. However, despite the considerable 
interest of the Langevin equation in the presence of the Lorentz force, a general relation 
analogous to Eq.~\eqref{VACT_BROWNIAN_2} and
derived from first principles, has so far been absent  \cite{chun2019effect, jimenez2006brownian, lemons1999brownian, pradhan2010nonexistence, kumar2009classical, baura2013study, jayannavar2007charged, abdoli2022tunable, abdoli2022escape} .

This paper fills that gap. Starting from the Langevin equation in the presence of an 
external magnetic field, we derive a general relation between the VACT and the FACT that 
holds for arbitrary conservative interactions. The derivation makes explicit how the broken 
time-reversal symmetry modifies the structure of the relation, and recovers the standard 
result Eq.~\eqref{VACT_BROWNIAN_2} in the zero-field limit. In the overdamped limit, the 
formalism provides a unified framework that encompasses and extends earlier results obtained 
for specific models~\cite{shinde2022strongly}, and clarifies the mechanism by which the 
Lorentz force enhances self-diffusion in interacting colloidal suspensions. As a concrete 
validation, we apply the general relation to a solvable model -- a dimer of charged 
particles in an external magnetic field -- reproducing and unifying known specific 
results~\cite{shinde2022strongly}.

The paper is organized as follows. In Sec.~I we analyze the time-symmetry structure of the 
Langevin equation in the presence of an external magnetic field, and discuss the 
consequences for the tensor structure of the VACT and FACT, following and extending the 
framework of Refs.~\cite{Coretti1, Coretti2}. In Sec.~II we derive the general VACT--FACT 
relation from the Langevin equation using these symmetry arguments. In Sec.~III we take the 
overdamped limit and obtain the analog of Eq.~\eqref{VACT_BROWNIAN_2} for systems subject 
to a Lorentz force. Finally, in Sec.~IV we apply the formalism to the dimer model, 
validating the general framework on a solvable case.

\section{Time Symmetries of the Equations of Motion in a Magnetic Field}

In this section we establish the symmetry operations available to the equations of motion 
in the presence of an external magnetic field, and derive the constraints they impose on 
the structure of the velocity and force autocorrelation tensors. We follow and extend the 
analysis of Coretti et al.~\cite{Coretti1, Coretti2}, which was carried out for 
Hamiltonian (noise-free, underdamped) dynamics. At the end of the section we show that 
the same symmetry structure carries over to the Langevin equation, which is the starting 
point for the remainder of this paper.

\subsection{Hamiltonian dynamics}

We consider the motion of a single charged particle of mass $m$ and charge $q$ in three 
dimensions, $\boldsymbol{x} = (x, y, z)$, subject to an external magnetic field 
$\mathbf{B} = B\hat{\mathbf{z}}$ and a position-dependent force $\boldsymbol{f} = 
-\nabla_{\boldsymbol{x}} U$, where $U$ may include both interparticle interactions and 
external confinement. The equation of motion reads
\begin{equation}\label{EOM_B}
    m\dot{\boldsymbol{v}}(t) = q\boldsymbol{v}(t)\times\mathbf{B} + \boldsymbol{f}(t).
\end{equation}
The state of the particle is described by the phase space supervector 
$\boldsymbol{X} = (x, y, z, v_x, v_y, v_z)$, whose time evolution is generated by the 
Liouville operator
\begin{equation}
    \boldsymbol{X}(t) = e^{t\Liou{B}}\boldsymbol{X}, \qquad
    \Liou{B}[\,\cdot\,] = \boldsymbol{v}\cdot\nabla_{\boldsymbol{x}}[\,\cdot\,] 
    + \frac{1}{m}\bigl(q\boldsymbol{v}\times\mathbf{B} 
    + \boldsymbol{f}\bigr)\cdot\nabla_{\boldsymbol{v}}[\,\cdot\,].
\end{equation}

\subsection{Time-reversal symmetry and its modification by the magnetic field}

In the absence of a magnetic field, the equations of motion are invariant under time 
reversal $t \to -t$ accompanied by $\boldsymbol{v} \to -\boldsymbol{v}$. In the presence 
of a magnetic field, however, the Lorentz force $q\boldsymbol{v}\times\mathbf{B}$ changes 
sign under this operation, so that invariance is restored only if one simultaneously 
reverses the field, $\mathbf{B} \to -\mathbf{B}$. We encode this combined transformation 
in the operator $\Top$, defined by its action on the supervector:
\begin{equation}
    \Top\boldsymbol{X} = (x, y, z, -v_x, -v_y, -v_z).
\end{equation}
The operator $\Top$ acts on a phase space function $\phi(\boldsymbol{X})$ by 
$\Top\phi(\boldsymbol{X}) = \phi(\Top\boldsymbol{X}) = \eta_\phi\,\phi(\boldsymbol{X})$, 
where the signature $\eta_\phi = \pm 1$ depends on whether $\phi$ is even or odd under 
velocity reversal. Furthermore, $\Top$ is Hermitian with respect to the inner product of 
phase space functions,
\begin{equation}\label{hermitian}
    \int d\boldsymbol{X}\; \phi(\boldsymbol{X})\,\Top\psi(\boldsymbol{X}) 
    = \int d\boldsymbol{X}\; \psi(\boldsymbol{X})\,\Top\phi(\boldsymbol{X}),
\end{equation}
which follows from the substitution $\boldsymbol{X} \to \Top\boldsymbol{X}$ in the 
integral. The invariance of Eq.~\eqref{EOM_B} under 
$(t,\boldsymbol{v},\mathbf{B})\to(-t,-\boldsymbol{v},-\mathbf{B})$ implies the 
phase space identity
\begin{equation}\label{time_symmetry}
    e^{t\Liou{B}}\boldsymbol{X} = \Top\, e^{-t\Liou{-B}}\,\Top\boldsymbol{X},
\end{equation}
and the same relation holds for any phase space function $\phi$,
\begin{equation}\label{time_symmetry_phi}
    e^{t\Liou{B}}\phi(\boldsymbol{X}) 
    = \Top\, e^{-t\Liou{-B}}\,\Top\phi(\boldsymbol{X}).
\end{equation}
The equilibrium time correlation function of two observables $\phi$ and $\psi$ at 
magnetic field $B$ is
\begin{equation}\label{Corr_fun}
    \bigl\langle\phi(t)\psi(0)\bigr\rangle\atB 
    = \int d\boldsymbol{X}\; \psi(\boldsymbol{X})\,
    e^{t\Liou{B}}\phi(\boldsymbol{X})\,\rho(\boldsymbol{X}),
\end{equation}
where $\rho(\boldsymbol{X})$ is the Boltzmann equilibrium distribution. We note that 
$\rho$ is independent of $B$, as required by the Bohr--Van Leeuwen theorem [CITE]: the 
equilibrium distribution of a charged particle in a magnetic field is identical to the 
field-free one, so that static averages carry no field dependence. Dynamic correlations, 
by contrast, do depend on $B$. Substituting Eq.~\eqref{time_symmetry_phi} into 
Eq.~\eqref{Corr_fun} and using the Hermiticity of $\Top$ together with stationarity, 
one obtains
\begin{equation}\label{TRS_1}
    \bigl\langle\phi(t)\psi(0)\bigr\rangle\atB 
    = \eta_\phi\,\eta_\psi\,
    \bigl\langle\psi(t)\phi(0)\bigr\rangle\atmB,
\end{equation}
which generalizes the standard time-reversal relation to finite $B$. In the absence of 
a magnetic field, Eq.~\eqref{TRS_1} reduces to the familiar result
$\langle\phi(t)\psi(0)\rangle = \eta_\phi\eta_\psi\langle\psi(t)\phi(0)\rangle$.
The price paid at finite $B$ is the field reversal on the right-hand side: 
Eq.~\eqref{TRS_1} alone does not constrain the correlation function at fixed $B$.

\subsection{Additional discrete symmetries for radially symmetric interactions}

To obtain constraints on correlation functions at fixed $B$, one needs additional 
symmetries that leave the magnetic field unchanged. Coretti et al.~\cite{Coretti1,Coretti2} 
identified four such transformations, valid when the potential $U$ is radially symmetric. 
They are spatial reflections combined with appropriate velocity sign changes, chosen so 
that the Lorentz force term in Eq.~\eqref{EOM_B} is preserved:
\begin{equation}\label{reflections}
\begin{alignedat}{2}
    \Rop{1}\boldsymbol{X} &= (-x,\phantom{-}y,\phantom{-}z,\phantom{-}v_x,-v_y,-v_z), 
    &\quad &\text{(reflection in }x\text{)}\\
    \Rop{2}\boldsymbol{X} &= (\phantom{-}x,-y,\phantom{-}z,-v_x,\phantom{-}v_y,-v_z), 
    &\quad &\text{(reflection in }y\text{)}\\
    \Rop{3}\boldsymbol{X} &= (\phantom{-}x,-y,-z,-v_x,\phantom{-}v_y,\phantom{-}v_z), 
    &\quad &\text{(reflection in }yz\text{)}\\
    \Rop{4}\boldsymbol{X} &= (-x,\phantom{-}y,-z,\phantom{-}v_x,-v_y,\phantom{-}v_z). 
    &\quad &\text{(reflection in }xz\text{)}
\end{alignedat}
\end{equation}
Each $\Rop{j}$ leaves Eq.~\eqref{EOM_B} invariant at fixed $B$, leading to the identity
\begin{equation}\label{TRS_2}
    e^{t\Liou{B}}\boldsymbol{X} 
    = \Rop{j}\,e^{-t\Liou{B}}\,\Rop{j}\boldsymbol{X}, 
    \qquad j = 1,\ldots,4,
\end{equation}
and consequently to the fixed-$B$ symmetry relations for correlation functions,
\begin{equation}\label{TRS_fixed_B}
    \bigl\langle\phi(t)\psi(0)\bigr\rangle\atB 
    = \eta^j_\phi\,\eta^j_\psi\,
    \bigl\langle\psi(t)\phi(0)\bigr\rangle\atB,
\end{equation}
where $\eta^j_\phi,\eta^j_\psi = \pm 1$ are the signatures of $\phi$ and $\psi$ under 
$\Rop{j}$.

\subsection{Tensor structure of the VACT and FACT}

Applying Eqs.~\eqref{TRS_1} and~\eqref{TRS_fixed_B} to the velocity and force 
components, one obtains the tensor structure of the VACT, 
$\Cvv\atB(t) = \langle\boldsymbol{v}(t)\otimes\boldsymbol{v}(0)\rangle\atB$, and of 
the FACT, $\Cff\atB(t) = \langle\boldsymbol{f}(t)\otimes\boldsymbol{f}(0)\rangle\atB$. 
With $\mathbf{B}\|\hat{\mathbf{z}}$, the magnetic field singles out the $z$-axis, 
reducing the symmetry to rotations in the $xy$-plane. As a result, couplings between 
the $z$-direction and the $xy$-plane vanish, the $xx$ and $yy$ components are equal, 
and an antisymmetric off-diagonal component in the $xy$-plane is permitted. Explicitly:
\begin{equation}\label{Structure_C}
\begin{alignedat}{3}
    &[\Cvv\atB]^{xz} = -[\Cvv\atB]^{zx} = 0, 
    &\qquad &[\Cff\atB]^{xz} = -[\Cff\atB]^{zx} = 0,\\
    &[\Cvv\atB]^{yz} = -[\Cvv\atB]^{zy} = 0, 
    &\qquad &[\Cff\atB]^{yz} = -[\Cff\atB]^{zy} = 0,\\
    &[\Cvv\atB]^{xy} = -[\Cvv\atB]^{yx}, 
    &\qquad &[\Cff\atB]^{xy} = -[\Cff\atB]^{yx},\\
    &[\Cvv\atB]^{xx} = [\Cvv\atB]^{yy}, 
    &\qquad &[\Cff\atB]^{xx} = [\Cff\atB]^{yy}.
\end{alignedat}
\end{equation}
The antisymmetric off-diagonal elements are odd functions of time. As a consequence, 
transposition of either tensor is equivalent to time reversal:
\begin{equation}\label{transpose}
    \bigl[\Cvv\atB\bigr]^\dagger(t) = \Cvv\atB(-t), 
    \qquad 
    \bigl[\Cff\atB\bigr]^\dagger(t) = \Cff\atB(-t).
\end{equation}
Combining Eq.~\eqref{transpose} with Eq.~\eqref{TRS_1} further gives
\begin{equation}\label{transpose_B}
    \bigl[\Cvv\atB\bigr]^\dagger(t) = \Cvv\atmB(t), 
    \qquad 
    \bigl[\Cff\atB\bigr]^\dagger(t) = \Cff\atmB(t),
\end{equation}
so that transposition is also equivalent to field reversal. In the absence of a magnetic 
field and for a radially symmetric potential, both tensors reduce to scalar multiples 
of the identity.

\subsection{Extension to the Langevin equation}

The analysis above was carried out for deterministic Hamiltonian dynamics. We now 
argue that the same symmetry structure carries over to the Langevin equation, which 
governs the stochastic dynamics of a particle coupled to a thermal bath:
\begin{equation}\label{langevin_B}
    m\dot{\boldsymbol{v}}(t) = -\gamma\boldsymbol{v}(t) 
    + q\boldsymbol{v}(t)\times\mathbf{B} 
    + \boldsymbol{f}(t) + \boldsymbol{\eta}(t),
\end{equation}
where $\gamma$ is the friction coefficient and $\boldsymbol{\eta}(t)$ is a Gaussian 
white noise with correlator
$\langle\boldsymbol{\eta}(t)\otimes\boldsymbol{\eta}(0)\rangle 
= 2\gamma k_BT\,\delta(t)\,\boldsymbol{I}$.
The noise term is isotropic and its correlator is independent of $B$, so it does not 
break any of the discrete symmetries identified above. The friction term $-\gamma\boldsymbol{v}$ 
is odd under velocity reversal and even under each of the spatial reflections $\Rop{j}$, 
consistent with the same signatures as the inertial term. Therefore, the tensor 
structures in Eqs.~\eqref{Structure_C}--\eqref{transpose_B} hold unchanged for the 
stochastic dynamics governed by Eq.~\eqref{langevin_B}, and we will make use of them 
throughout the remainder of this paper.

\section{Relation between the velocity and force autocorrelation tensors}

In this section we derive a relation expressing the VACT in terms of the FACT, starting 
from the Langevin equation and using the time-symmetry results established above. We 
first obtain the relation for the full, underdamped Langevin dynamics, and then 
specialize it to the overdamped (Brownian) limit, where it takes a closed algebraic form. 
Finally we discuss its consequence for the self-diffusion of the particle.

\subsection{Underdamped Langevin dynamics}

The Langevin equation~\eqref{langevin_B} admits the integral representation
\begin{equation}\label{langevin_B_int}
    \boldsymbol{v}(t) = e^{-\mathbf{G}t}\boldsymbol{v}(0) 
    + \frac{1}{m}\int_0^t d\tau\; e^{-\mathbf{G}(t-\tau)}
    \bigl[\boldsymbol{f}(\tau) + \boldsymbol{\eta}(\tau)\bigr],
\end{equation}
where the combined effect of friction and Lorentz force is encoded in the matrix
\begin{equation}
    \mathbf{G} = 
    \begin{bmatrix}
        1/\tau_{th} & -\omega_c & 0 \\
        \omega_c & 1/\tau_{th} & 0 \\
        0 & 0 & 1/\tau_{th}
    \end{bmatrix}, 
    \qquad
    \omega_c = \frac{qB}{m}, \qquad \tau_{th} = \frac{m}{\gamma}.
\end{equation}
The matrix $\mathbf{G}$ depends on two independent parameters: the cyclotron frequency 
$\omega_c$ and the thermalization time $\tau_{th}$. Their ratio defines the dimensionless 
parameter $\kappa = \omega_c\tau_{th}$ already introduced in the Introduction.

The exponential $e^{-\mathbf{G}t}$ is readily evaluated, and is directly related to the 
velocity autocorrelation tensor of a free particle (i.e.\ in the absence of the 
interaction force, $\boldsymbol{f} = 0$),
\begin{equation}\label{free_kernel}
    \Ctilde\atB(t) = e^{-\mathbf{G}t}\bigl\langle\boldsymbol{v}(0)\otimes\boldsymbol{v}(0)\bigr\rangle 
    = \frac{k_BT}{\gamma}\frac{e^{-t/\tau_{th}}}{\tau_{th}}
    \begin{bmatrix}
        \cos(\omega_c t) & \sin(\omega_c t) & 0 \\
        -\sin(\omega_c t) & \cos(\omega_c t) & 0 \\
        0 & 0 & 1
    \end{bmatrix}.
\end{equation}
This free-particle kernel satisfies all the symmetries of correlation functions 
established in the previous section. We note that its $zz$ component is independent of 
the magnetic field; this is a feature of the free particle and does not hold in general, 
as it depends on the specific form of the interaction force~\cite{Bonellat2013}.

To derive the relation between the VACT and the FACT, we take the outer product of 
Eq.~\eqref{langevin_B} with $\boldsymbol{v}(0)$. Using the causality condition 
$\langle\boldsymbol{\eta}(t)\otimes\boldsymbol{v}(0)\rangle = 0$, we obtain
\begin{equation}\label{VACT_FVCT}
    \Cvv\atB(t) = \Ctilde\atB(t) 
    + \beta\int_0^t d\tau\; \Ctilde\atB(t-\tau)
    \bigl\langle\boldsymbol{f}(\tau)\otimes\boldsymbol{v}(0)\bigr\rangle\atB, 
    \qquad \beta = (k_BT)^{-1}.
\end{equation}
To evaluate the force--velocity cross-correlator appearing here, we take the outer 
product of Eq.~\eqref{langevin_B_int} with $\boldsymbol{f}(0)$. The term depending on the 
initial velocity vanishes, since velocity and position-dependent variables are 
uncorrelated in equilibrium, $\langle\boldsymbol{v}(0)\otimes\boldsymbol{f}(0)\rangle = 0$, 
leaving
\begin{equation}\label{FVCT_FACT}
    \bigl\langle\boldsymbol{v}(t)\otimes\boldsymbol{f}(0)\bigr\rangle\atB 
    = \beta\int_0^t d\tau\; \Ctilde\atB(t-\tau)\,\Cff\atB(\tau).
\end{equation}
We now use the time-reversal relation~\eqref{TRS_1} to exchange the time arguments at 
the cost of reversing the field,
\begin{equation}\label{FVCT_FACT_1}
    \bigl\langle\boldsymbol{v}(0)\otimes\boldsymbol{f}(t)\bigr\rangle\atB 
    = -\bigl\langle\boldsymbol{v}(t)\otimes\boldsymbol{f}(0)\bigr\rangle\atmB 
    = -\beta\int_0^t d\tau\; \Ctilde\atmB(t-\tau)\,\Cff\atmB(\tau),
\end{equation}
and the transpose relation~\eqref{transpose_B} to convert the field reversal back into a 
transposition at fixed $B$,
\begin{equation}\label{FVCT_FACT_2}
    \bigl\langle\boldsymbol{v}(0)\otimes\boldsymbol{f}(t)\bigr\rangle\atB 
    = -\beta\int_0^t d\tau\; \Ctilde\atB^\dagger(t-\tau)\,\Cff\atB^\dagger(\tau).
\end{equation}
Inserting the transpose of Eq.~\eqref{FVCT_FACT_2} into Eq.~\eqref{VACT_FVCT}, we arrive 
at the first main result of this work: an expression for the VACT that depends only on 
the FACT,
\begin{equation}\label{VACT_FACT}
    \Cvv\atB(t) = \Ctilde\atB(t) 
    - \beta^2\int_0^t d\tau\; \Ctilde\atB(t-\tau)
    \int_0^\tau d\tau_1\; \Cff\atB(\tau_1)\,\Ctilde\atB(\tau-\tau_1).
\end{equation}
The second term is a double convolution of the FACT with two copies of the free-particle 
kernel, depending on $t-\tau$ and $\tau-\tau_1$ respectively. Since both kernels decay 
exponentially on the timescale $\tau_{th}$, the integral effectively confines the FACT to 
a region of size $\tau_{th}\times\tau_{th}$ around $t$. In the absence of a magnetic 
field, Eq.~\eqref{VACT_FACT} reduces to
\begin{equation}\label{VACT_FACT_ORDINARY}
    \boldsymbol{C}^{vv}(t) = \boldsymbol{I}\,\frac{k_BT}{\gamma}\frac{e^{-t/\tau_{th}}}{\tau_{th}} 
    - \frac{1}{\gamma^2}\int_0^t d\tau\; \frac{1}{\tau_{th}}e^{-(t-\tau)/\tau_{th}}
    \int_0^\tau d\tau_1\; \frac{1}{\tau_{th}}e^{-(\tau-\tau_1)/\tau_{th}}\,\boldsymbol{C}^{ff}(\tau_1).
\end{equation}

\subsection{Overdamped limit}

Brownian dynamics emerges as an accurate approximation of the Langevin equation when the 
thermalization time is much shorter than the decay time of the force autocorrelation. 
This separation of timescales is implemented by taking the limit $\tau_{th}\to 0$. 
Applied to Eq.~\eqref{VACT_FACT_ORDINARY}, this limit directly reproduces 
Eq.~\eqref{VACT_BROWNIAN_2}. We now generalize this result to finite magnetic field.

Taking the limit $\tau_{th}\to 0$ of Eq.~\eqref{VACT_FACT} at fixed $\kappa$ requires care, 
since the free-particle kernel $\Ctilde\atB(t)$ becomes sharply peaked at $t=0$. Following 
Chun et al.~\cite{Non_white_noise}, the kernel reduces, for finite $\kappa$, to a 
combination of one-sided Dirac delta distributions,
\begin{equation}\label{kernel}
    \lim_{\tau_{th}\to 0}\Ctilde\atB(t) \simeq \delta^+(t)\,\boldsymbol{D} + \delta^-(t)\,\boldsymbol{D}^\dagger,
    \qquad
    \boldsymbol{D} = \frac{k_BT}{\gamma(1+\kappa^2)}
    \begin{bmatrix}
        1 & \kappa & 0 \\
        -\kappa & 1 & 0 \\
        0 & 0 & 1+\kappa^2
    \end{bmatrix},
\end{equation}
where the one-sided deltas satisfy
\begin{subequations}
\begin{align}
    \int_0^{t>0} d\tau\; \delta^+(\tau)f(\tau) &= \lim_{\epsilon\to 0^+}f(\epsilon), 
    &\int_0^{t>0} d\tau\; \delta^-(\tau)f(\tau) &= 0, \\
    \int_0^{t<0} d\tau\; \delta^-(\tau)f(\tau) &= \lim_{\epsilon\to 0^-}f(\epsilon), 
    &\int_0^{t<0} d\tau\; \delta^+(\tau)f(\tau) &= 0.
\end{align}
\end{subequations}
Equation~\eqref{kernel} respects the symmetries~\eqref{transpose} and~\eqref{transpose_B}, 
and reduces, for $\kappa = 0$, to the ordinary result 
$\Ctilde(t) \simeq \delta(t)\,2k_BT/\gamma$.

Inserting Eq.~\eqref{kernel} into Eq.~\eqref{VACT_FACT} yields the second main result of 
this work: a closed relation between the VACT and the FACT for overdamped dynamics in the 
presence of a magnetic field,
\begin{equation}\label{VACT_FACT_OVER}
    \bigl\langle\boldsymbol{v}(t)\otimes\boldsymbol{v}(0)\bigr\rangle\atB 
    = \delta^+(t)\,\boldsymbol{D} 
    - \beta^2\,\boldsymbol{D}\,\bigl\langle\boldsymbol{f}(t)\otimes\boldsymbol{f}(0)\bigr\rangle\atB\,\boldsymbol{D}.
\end{equation}
For $\kappa = 0$, Eq.~\eqref{VACT_FACT_OVER} reduces to the ordinary result 
Eq.~\eqref{VACT_BROWNIAN_2}. Decomposing Eq.~\eqref{VACT_FACT_OVER} into components shows 
that the off-diagonal elements of the FACT feed into every component of the VACT -- a 
genuine peculiarity of the presence of the magnetic field, absent in the ordinary case:

\begin{equation}\label{diag_off}
\begin{split}
    [\Cvv\atB]^{xx}(t) &=
    \frac{k_BT}{\gamma(1+\kappa^2)}\delta^+(t) 
    - \frac{1}{\gamma^2(1+\kappa^2)^2}\Bigl[(1-\kappa^2)[\Cff\atB]^{xx}(t) - 2\kappa[\Cff\atB]^{xy}(t)\Bigr], \\
    [\Cvv\atB]^{xy}(t)  &= 
    \kappa\frac{k_BT}{\gamma(1+\kappa^2)}\delta^+(t) 
    - \frac{1}{\gamma^2(1+\kappa^2)^2}\Bigl[(1-\kappa^2)[\Cff\atB]^{xy}(t) + 2\kappa[\Cff\atB]^{xx}(t)\Bigr], \\
    [\Cvv\atB]^{zz}(t) &= \frac{k_BT}{\gamma}\delta^+(t) - \frac{1}{\gamma^2}[\Cff\atB]^{zz}(t). 
\end{split}
\end{equation}

The other elements of the tensor follows from the symmetry relations of Eqs.~(\ref{Structure_C}).
It can be observed that the magnetic field, through $\kappa$, multiplies the FACT components with odd or even 
powers depending on their odd or even character in time, in a way that preserves the 
time-reversal properties of the VACT components on the left-hand side.

The overdamped equation of motion consistent with Eq.~\eqref{VACT_FACT_OVER} follows 
directly from inserting the kernel~\eqref{kernel} into Eq.~\eqref{langevin_B_int}, as 
already obtained in Ref.~\cite{Non_white_noise},
\begin{equation}\label{Brownian_B}
    \boldsymbol{v}(t) = \beta\boldsymbol{D}\boldsymbol{f}(t) + \boldsymbol{\xi}(t),
\end{equation}
where the autocorrelation of the effective noise $\boldsymbol{\xi}(t)$ follows from 
Eq.~\eqref{VACT_FACT_OVER} upon setting $\boldsymbol{f} = 0$,
\begin{equation}
    \bigl\langle\boldsymbol{\xi}(t)\otimes\boldsymbol{\xi}(0)\bigr\rangle 
    = \delta^+(t)\,\boldsymbol{D} + \delta^-(t)\,\boldsymbol{D}^\dagger.
\end{equation}
Equation~\eqref{VACT_FACT_OVER} further implies a relation between the noise--force 
cross-correlator and the FACT, generalizing the result of Ref.~\cite{Akesson1985} 
obtained for Brownian dynamics without a magnetic field. Starting from the transpose of 
Eq.~\eqref{FVCT_FACT_2} and taking the limit $\tau_{th}\to 0$, we obtain
\begin{equation}\label{FORCE_NOISE_MF_1}
    \bigl\langle\boldsymbol{f}(t)\otimes\boldsymbol{v}(0)\bigr\rangle\atB 
    = -\beta\,\bigl\langle\boldsymbol{f}(t)\otimes\boldsymbol{f}(0)\bigr\rangle\atB\,\boldsymbol{D}.
\end{equation}
Substituting Eq.~\eqref{Brownian_B} into Eq.~\eqref{FORCE_NOISE_MF_1} gives the 
noise--force correlator in terms of the FACT,
\begin{equation}\label{FORCE_NOISE_MF_2}
    \bigl\langle\boldsymbol{f}(t)\otimes\boldsymbol{\xi}(0)\bigr\rangle\atB 
    = -\beta\,\bigl\langle\boldsymbol{f}(t)\otimes\boldsymbol{f}(0)\bigr\rangle\atB\,
    \bigl(\boldsymbol{D} + \boldsymbol{D}^\dagger\bigr).
\end{equation}
This is the third main result of this work. In the absence of a magnetic field, $\kappa = 0$, 
it reduces to the result of Ref.~\cite{Akesson1985},
\begin{equation}\label{FORCE_NOISE_2}
    \gamma\,\bigl\langle\boldsymbol{f}(t)\otimes\boldsymbol{\xi}(0)\bigr\rangle 
    = -2\,\bigl\langle\boldsymbol{f}(t)\otimes\boldsymbol{f}(0)\bigr\rangle.
\end{equation}

\subsection{Self-diffusion}

The mean square displacement (MSD) of the particle follows from the double time integral 
of the trace of the VACT,
\begin{equation}
    \bigl\langle|\Delta\boldsymbol{R}|^2(t)\bigr\rangle 
    = \int_0^t d\tau\int_0^t d\tau_1\; 
    \mathrm{Tr}\bigl[\bigl\langle\boldsymbol{v}(\tau)\otimes\boldsymbol{v}(\tau_1)\bigr\rangle\atB\bigr].
\end{equation}
Using stationarity, 
$\mathrm{Tr}[\langle\boldsymbol{v}(\tau)\otimes\boldsymbol{v}(\tau_1)\rangle\atB] 
= \mathrm{Tr}[\langle\boldsymbol{v}(\tau-\tau_1)\otimes\boldsymbol{v}(0)\rangle\atB]$, 
together with the even character of the trace of the VACT in time, the double integral 
simplifies to
\begin{equation}\label{DR_0}
    \bigl\langle|\Delta\boldsymbol{R}|^2(t)\bigr\rangle 
    = 2\int_0^t d\tau\,(t-\tau)\,\mathrm{Tr}\bigl[\Cvv\atB(\tau)\bigr].
\end{equation}
Inserting Eq.~\eqref{diag_off} into Eq.~\eqref{DR_0} yields an expression for the MSD that 
depends explicitly on the off-diagonal elements of the FACT,
\begin{equation}\label{DR}
\begin{split}
    \frac{1}{2}\bigl\langle|\Delta\boldsymbol{R}|^2(t)\bigr\rangle 
    = \frac{k_BT}{\gamma}\frac{3+\kappa^2}{1+\kappa^2}\,t 
    &- \frac{1}{\gamma^2}\int_0^t d\tau\,(t-\tau)\,[\Cff\atB]^{zz}(\tau) \\
    &- \frac{2}{\gamma^2(1+\kappa^2)^2}\int_0^t d\tau\,(t-\tau)
    \Bigl[(1-\kappa^2)[\Cff\atB]^{xx}(\tau) - 2\kappa[\Cff\atB]^{xy}(\tau)\Bigr].
\end{split}
\end{equation}
Equation~\eqref{DR} shows that, in contrast to the field-free case, the self-diffusion of 
the particle is sensitive not only to the diagonal elements of the FACT but also to its 
off-diagonal, antisymmetric part, weighted by $\kappa$. This is precisely the mechanism 
through which oscillations in the FACT of interacting odd systems -- such as those 
reported for hard-sphere particles in Ref.~\cite{Osc_Forc_Autcorr} -- translate into a measurable 
enhancement of self-diffusion, as we discuss further in Sec.~IV.

\section{Brownian motion of a dimer with the Lorentz force}

In this section we apply the formalism developed above to an exactly solvable model: two 
charged particles, labeled $1$ and $2$, interacting through a harmonic binding potential 
and subject to friction and a Lorentz force. This model both validates the general 
relations of Secs.~II and III and reveals, in closed form, how the relative charge of the 
two particles controls the qualitative behavior of the dimer. When the particles carry 
opposite charges, the Lorentz forces acting on them are opposite in sign and largely 
cancel in the relative coordinate, leaving only a transient, short-time signature of the 
magnetic field that disappears at long times. When the particles carry the same charge, 
this cancellation does not occur: the relative coordinate fully inherits the rotational 
coupling induced by the field, the FACT develops a genuine oscillatory behavior -- the 
exactly solvable realization of the phenomenon discussed in the Introduction -- and the 
in-plane self-diffusion is persistently suppressed at long times. These results are 
consistent with the previous findings of Refs.~\cite{shinde2022strongly, Osc_Forc_Autcorr}.

\subsection{Dimer with opposite charges}

We first consider oppositely charged particles. Since the Lorentz force changes sign with 
the particle charge, the two particles experience opposite cyclotron couplings, 
$\kappa_1 = -\kappa_2 \equiv \kappa$, and correspondingly $\boldsymbol{D} = \boldsymbol{D}_1 
= \boldsymbol{D}_2^{\dagger}$. The equations of motion read
\begin{subequations}
\begin{align}
    \dot{\boldsymbol{x}}_1(t) &= -\alpha\beta\boldsymbol{D}_1\boldsymbol{x}(t) + \boldsymbol{\eta}_1(t), \label{EOM_dimer_opposite_1} \\
    \dot{\boldsymbol{x}}_2(t) &= \phantom{-}\alpha\beta\boldsymbol{D}_2\boldsymbol{x}(t) + \boldsymbol{\eta}_2(t), \label{EOM_dimer_opposite_2}
\end{align}
\end{subequations}
where $\boldsymbol{x}(t) = \boldsymbol{x}_1(t) - \boldsymbol{x}_2(t)$ is the relative 
coordinate, $\alpha$ is the strength of the harmonic potential, and the noise correlators 
are
\begin{equation}
\begin{split}
    \bigl\langle\boldsymbol{\eta}_1(t)\otimes\boldsymbol{\eta}_1(0)\bigr\rangle 
    &= \bigl\langle\boldsymbol{\eta}_2(t)\otimes\boldsymbol{\eta}_2(0)\bigr\rangle^\dagger 
    = \delta^+(t)\boldsymbol{D} + \delta^-(t)\boldsymbol{D}^\dagger, \\
    \bigl\langle\boldsymbol{\eta}_1(t)\otimes\boldsymbol{\eta}_2(0)\bigr\rangle 
    &= \bigl\langle\boldsymbol{\eta}_2(t)\otimes\boldsymbol{\eta}_1(0)\bigr\rangle = 0.
\end{split}
\end{equation}
Subtracting Eq.~\eqref{EOM_dimer_opposite_2} from Eq.~\eqref{EOM_dimer_opposite_1}, the 
opposite signs of $\boldsymbol{D}_1$ and $\boldsymbol{D}_2$ in front of the velocity-rotating 
off-diagonal terms cancel, and only their common diagonal part survives in the relative 
dynamics:
\begin{equation}\label{dimer_same_charge_x}
    \dot{\boldsymbol{x}}(t) = -\frac{\boldsymbol{\Gamma}}{\tau}\boldsymbol{x}(t) 
    + \boldsymbol{\eta}_1(t) - \boldsymbol{\eta}_2(t),
\end{equation}
where we used $\boldsymbol{D}_1 + \boldsymbol{D}_2 = (k_BT/\gamma)\boldsymbol{\Gamma}$, with
\begin{equation}
    \boldsymbol{\Gamma} = 
    \begin{bmatrix}
        (1+\kappa^2)^{-1} & 0 & 0 \\
        0 & (1+\kappa^2)^{-1} & 0 \\
        0 & 0 & 1
    \end{bmatrix},
    \qquad \tau = \frac{\gamma}{2\alpha}.
\end{equation}
The diagonal structure of $\boldsymbol{\Gamma}$ is the direct consequence of the charge 
cancellation: no rotational (cyclotron-like) coupling survives in the relative coordinate. 
Equation~\eqref{dimer_same_charge_x} is formally solved by
\begin{equation}\label{dimer_same_charge_x_int}
    \boldsymbol{x}(t) = e^{-\boldsymbol{\Gamma}t/\tau}\boldsymbol{x}(0) 
    + \int_0^t dt_1\; e^{-\boldsymbol{\Gamma}(t-t_1)/\tau}
    \bigl(\boldsymbol{\eta}_1(t_1) - \boldsymbol{\eta}_2(t_1)\bigr),
\end{equation}
and the outer product with $\boldsymbol{x}(0)$ yields the force autocorrelation tensor,
\begin{equation}\label{dimer_FACF_opposite}
\begin{split}
    [\Cff\atB]^{xx}(t) &= \alpha^2\bigl\langle x(t)\,x(0)\bigr\rangle\atB 
    = \frac{\alpha}{\beta}\,e^{-\frac{t}{\tau}\frac{1}{1+\kappa^2}}, \\
    [\Cff\atB]^{xy}(t) &= \alpha^2\bigl\langle x(t)\,y(0)\bigr\rangle\atB = 0, \\
    [\Cff\atB]^{zz}(t) &= \alpha^2\bigl\langle z(t)\,z(0)\bigr\rangle\atB 
    = \frac{\alpha}{\beta}\,e^{-t/\tau}.
\end{split}
\end{equation}
As anticipated, the diagonal structure of $\boldsymbol{\Gamma}$ produces a purely 
exponential, non-oscillatory FACT: with no rotational coupling in the relative dynamics, 
there is no mechanism to generate oscillations.

\begin{figure}[h!]
\centering
\begin{subfigure}{0.45\textwidth}
    \includegraphics[width=\textwidth]{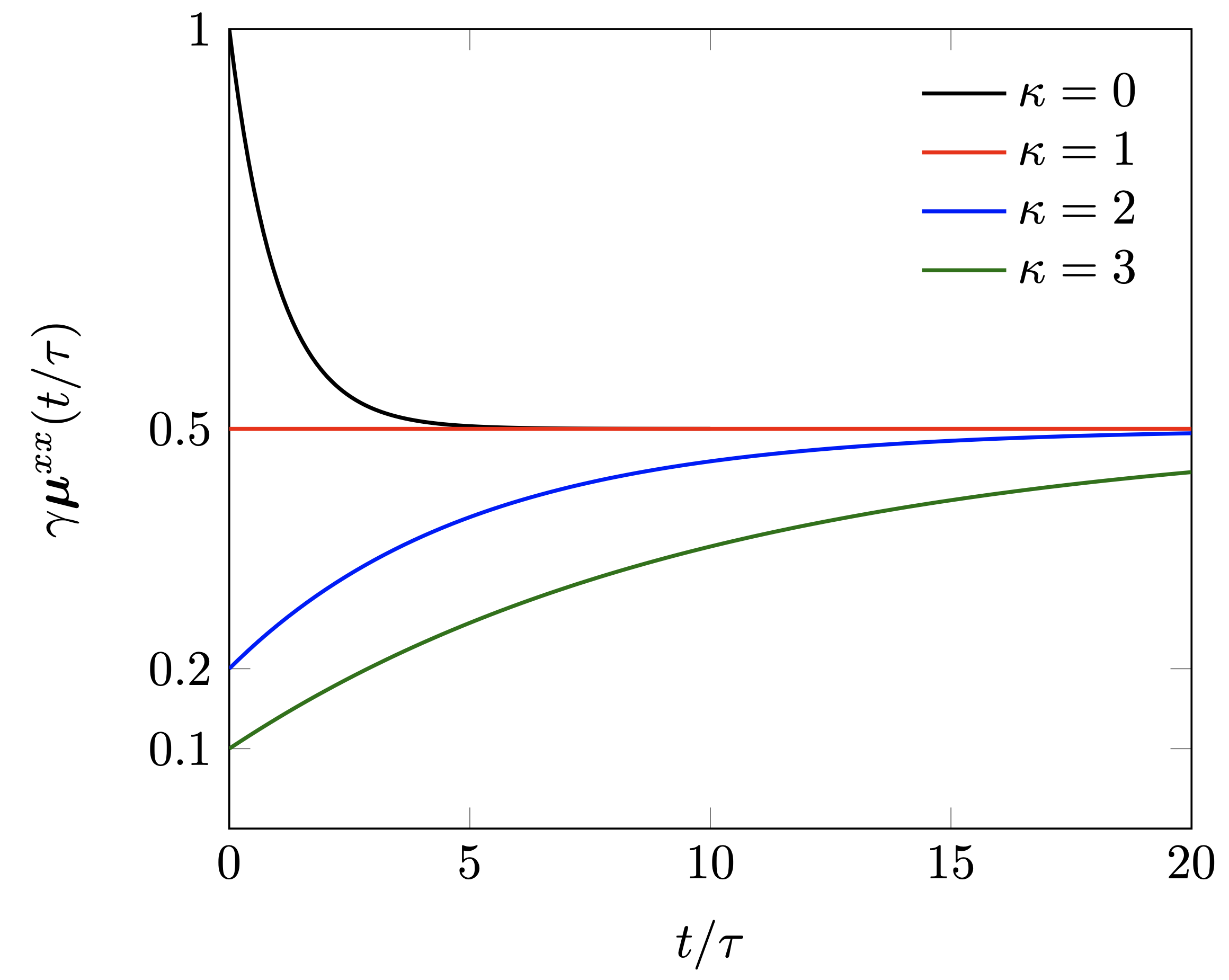}
    \caption{}
    \label{mobility_dimer_opposite}
\end{subfigure}
\hfill
\begin{subfigure}{0.45\textwidth}
    \includegraphics[width=\textwidth]{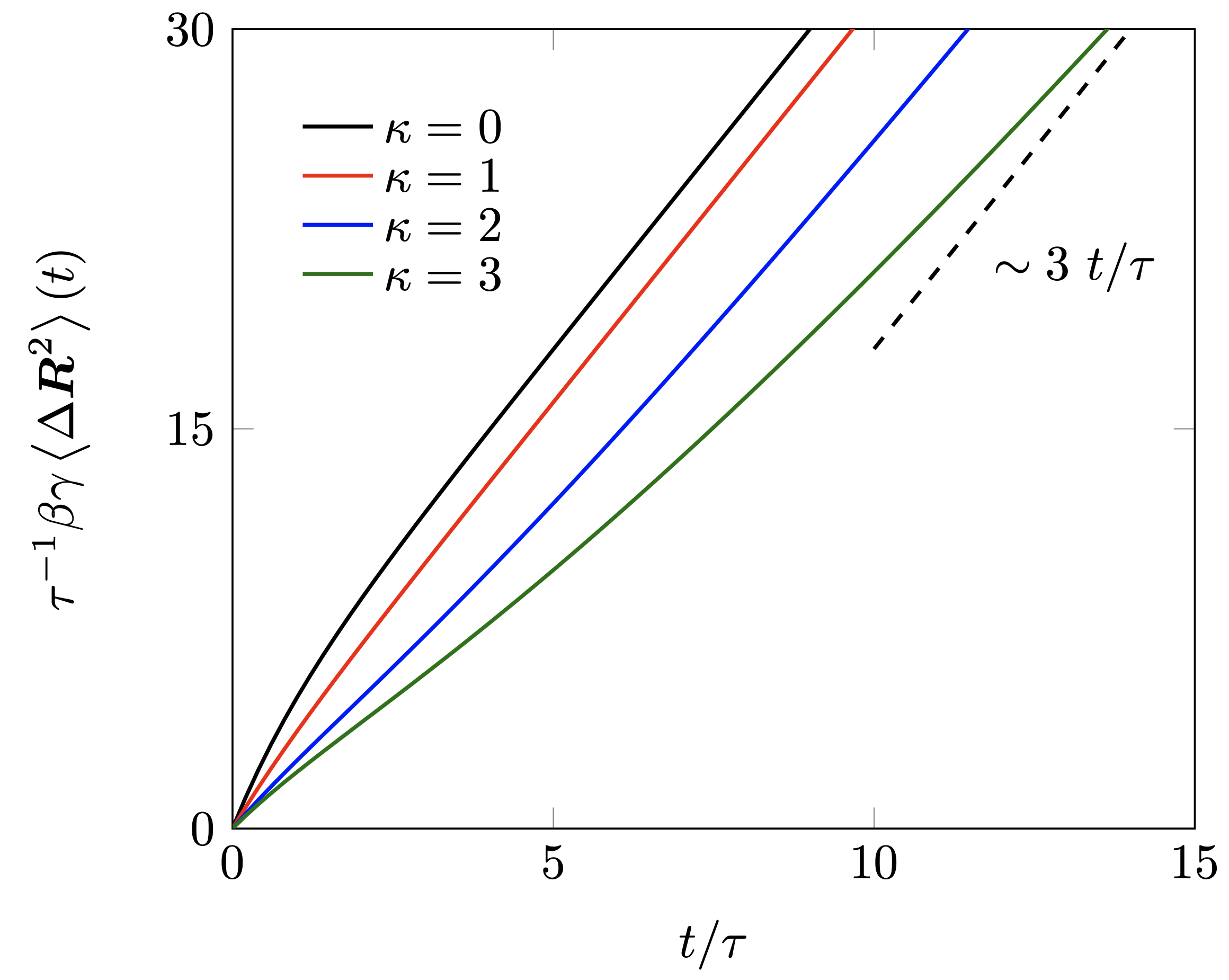}
    \caption{}
    \label{MSD_opposite}
\end{subfigure}    
\caption{(a) Plot of the diagonal component of the mobility tensor, $\gamma\boldsymbol{\mu}_{xx}$, as a function of the dimensionless time $t/\tau$ and magnetic field $\kappa$. (b) Plot of the mean square displacement, $\beta\gamma\tau^{-1}\left<\vert \Delta \boldsymbol R\vert^2_{(t)} \right>$ as a function of the dimensionless time $t/\tau$ and magnetic field $\kappa$.}
\label{fig}
\end{figure}

The time-dependent mobility tensor, $\boldsymbol{\mu}(t) = \beta\int_0^t d\tau\, 
\Cvv\atB(\tau)$, follows from Eq.~\eqref{VACT_FACT_OVER}:
\begin{equation}\label{D_DIMER_opposite}
\begin{split}
    \gamma\mu_{xx}(t) &= \frac{1}{1+\kappa^2} 
    - \frac{1}{2}\frac{1-\kappa^2}{1+\kappa^2}
    \left(1 - e^{-\frac{t}{\tau}\frac{1}{1+\kappa^2}}\right), \\
    \gamma\mu_{xy}(t) &= \frac{\kappa}{1+\kappa^2}, \\
    \gamma\mu_{zz}(t) &= \gamma\mu_{xx}(t)\big|_{\kappa=0}.
\end{split}
\end{equation}
The diagonal component $\mu_{xx}$ is shown as a function of $\kappa$ and $t/\tau$ in 
Fig.~\ref{mobility_dimer_opposite}. At short times, $\mu_{xx}(0) = 1/[\gamma(1+\kappa^2)]$ 
is suppressed by the magnetic field, exactly as for a free charged particle. At long 
times, however, $\mu_{xx}$ relaxes to $1/(2\gamma)$, \emph{independently of $\kappa$}: the 
field's suppressive effect is transient and is entirely washed out at long times, so the 
particle ultimately diffuses at the same rate it would in the absence of the field. This 
short-to-long-time recovery is the central feature of the opposite-charge dimer. A curious 
limiting case occurs at $\kappa = 1$, where the prefactor $(1-\kappa^2)$ vanishes and 
$\mu_{xx}$ becomes entirely time-independent.

The mean square displacement follows from Eq.~\eqref{DR}:
\begin{equation}\label{R_DIMER_opposite}
\begin{split}
    \bigl\langle|\Delta\boldsymbol{R}|^2(t)\bigr\rangle 
    = 6\frac{k_BT}{\gamma}t\Biggl\{
    &\frac{1+\kappa^2/3}{1+\kappa^2} 
    - \frac{1}{3}\frac{1-\kappa^2}{1+\kappa^2}
    \left[1 - \frac{1+\kappa^2}{t/\tau}\left(1-e^{-\frac{t}{\tau}\frac{1}{1+\kappa^2}}\right)\right] \\
    - &\frac{1}{6}\left(1 - \frac{1}{t/\tau}\left(1-e^{-t/\tau}\right)\right)
    \Biggr\},
\end{split}
\end{equation}
shown in Fig.~\ref{MSD_opposite}. Consistent with the mobility, the long-time MSD growth 
rate is independent of the magnetic field, $\sim 3(k_BT/\gamma)\,t$, while the short-time 
growth rate is suppressed by $\kappa$, $\sim (k_BT/\gamma)[(6+2\kappa^2)/(1+\kappa^2)]\,t$: 
the field slows the initial spreading of the dimer, but this suppression is recovered at 
long times.

\subsection{Dimer with same charges}

We now consider particles with equal charge, $\kappa_1 = \kappa_2 \equiv \kappa$, so that 
$\boldsymbol{D}_1 = \boldsymbol{D}_2 = \boldsymbol{D}$. In contrast to the opposite-charge 
case, the rotational coupling no longer cancels: both particles are deflected by the 
Lorentz force in the same sense, and the relative coordinate fully inherits this rotation. 
The equations of motion are
\begin{equation}\label{EOM_dimer}
\begin{split}
    \dot{\boldsymbol{x}}_1(t) &= -\alpha\beta\boldsymbol{D}\boldsymbol{x}(t) + \boldsymbol{\eta}_1(t), \\
    \dot{\boldsymbol{x}}_2(t) &= \phantom{-}\alpha\beta\boldsymbol{D}\boldsymbol{x}(t) + \boldsymbol{\eta}_2(t),
\end{split}
\end{equation}
with noise correlator $\langle\boldsymbol{\eta}_i(t)\otimes\boldsymbol{\eta}_j(0)\rangle 
= \delta_{ij}\bigl(\delta^+(t)\boldsymbol{D} + \delta^-(t)\boldsymbol{D}^\dagger\bigr)$. The 
relative coordinate evolves as
\begin{equation}\label{dimer_rel}
    \boldsymbol{x}(t) = e^{-2\alpha\beta\boldsymbol{D}t}\boldsymbol{x}(0) 
    + \int_0^t dt_1\; e^{-2\alpha\beta\boldsymbol{D}(t-t_1)}
    \bigl(\boldsymbol{\eta}_1(t_1) - \boldsymbol{\eta}_2(t_1)\bigr),
\end{equation}
and the outer product with $\boldsymbol{x}(0)$ gives the force autocorrelation tensor,
\begin{equation}\label{dimer_FACF}
\begin{split}
    [\Cff\atB]^{xx}(t) &= \frac{\alpha}{\beta}\,e^{-t/\tau_\kappa}\cos\!\left(\frac{\kappa t}{\tau_\kappa}\right), \\
    [\Cff\atB]^{xy}(t) &= \frac{\alpha}{\beta}\,e^{-t/\tau_\kappa}\sin\!\left(\frac{\kappa t}{\tau_\kappa}\right), \\
    [\Cff\atB]^{zz}(t) &= \frac{\alpha}{\beta}\,e^{-t/\tau},
\end{split}
\end{equation}
with $\tau_\kappa = \tau(1+\kappa^2)$. Unlike the opposite-charge case, the magnetic field 
now generates a genuinely oscillatory FACT: the off-diagonal coupling in $\boldsymbol{D}$ 
survives in the relative dynamics and rotates the force correlations as they decay. This 
is the exactly solvable, minimal realization of the oscillatory-FACT phenomenon discussed 
in the Introduction for interacting odd systems~\cite{D_S_C_P}.

\begin{figure}[h!]
\centering
\begin{subfigure}{0.45\textwidth}
    \includegraphics[width=\textwidth]{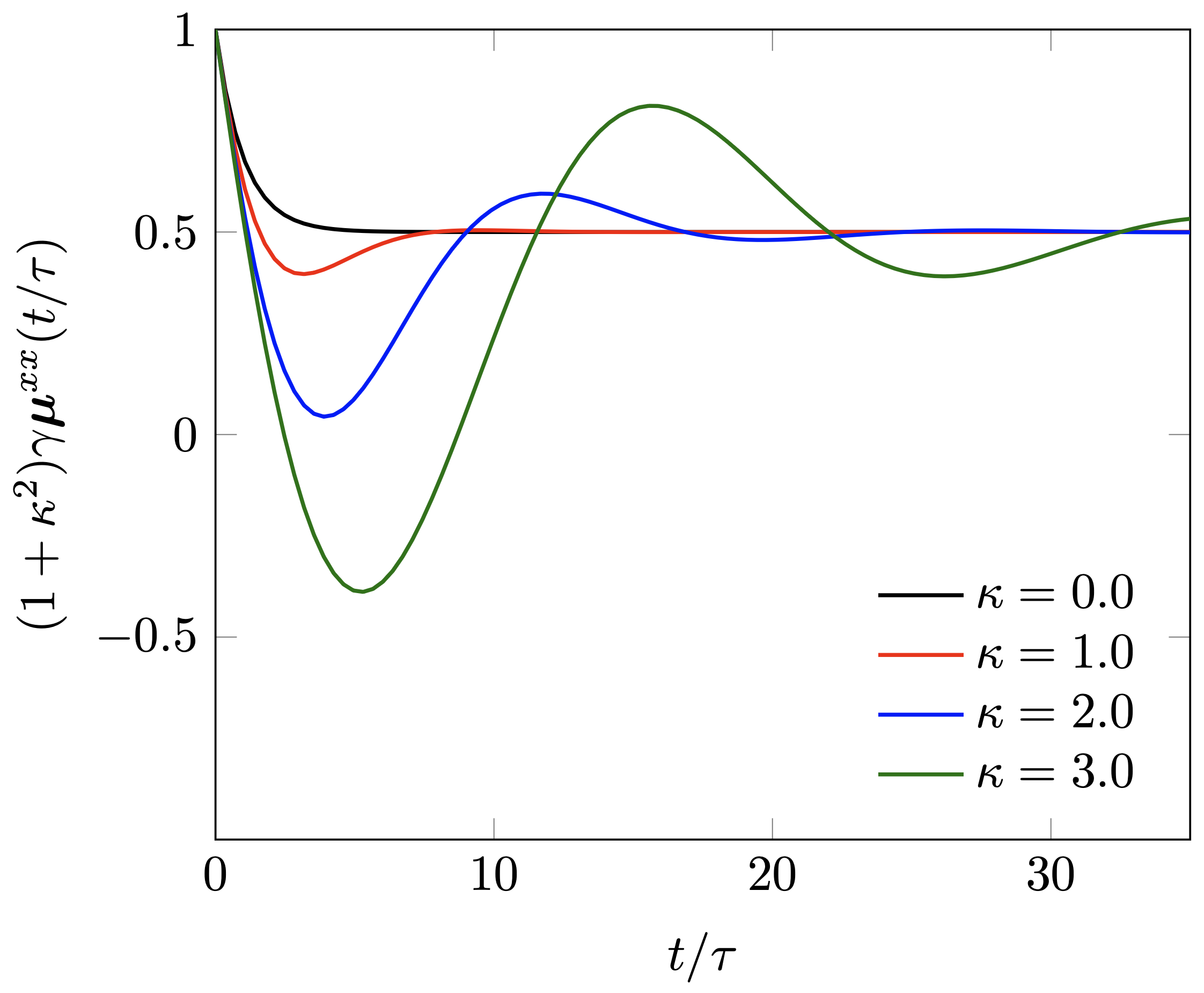}
    \caption{}
    \label{fig:first}
\end{subfigure}
\hfill
\begin{subfigure}{0.47\textwidth}
    \includegraphics[width=\textwidth]{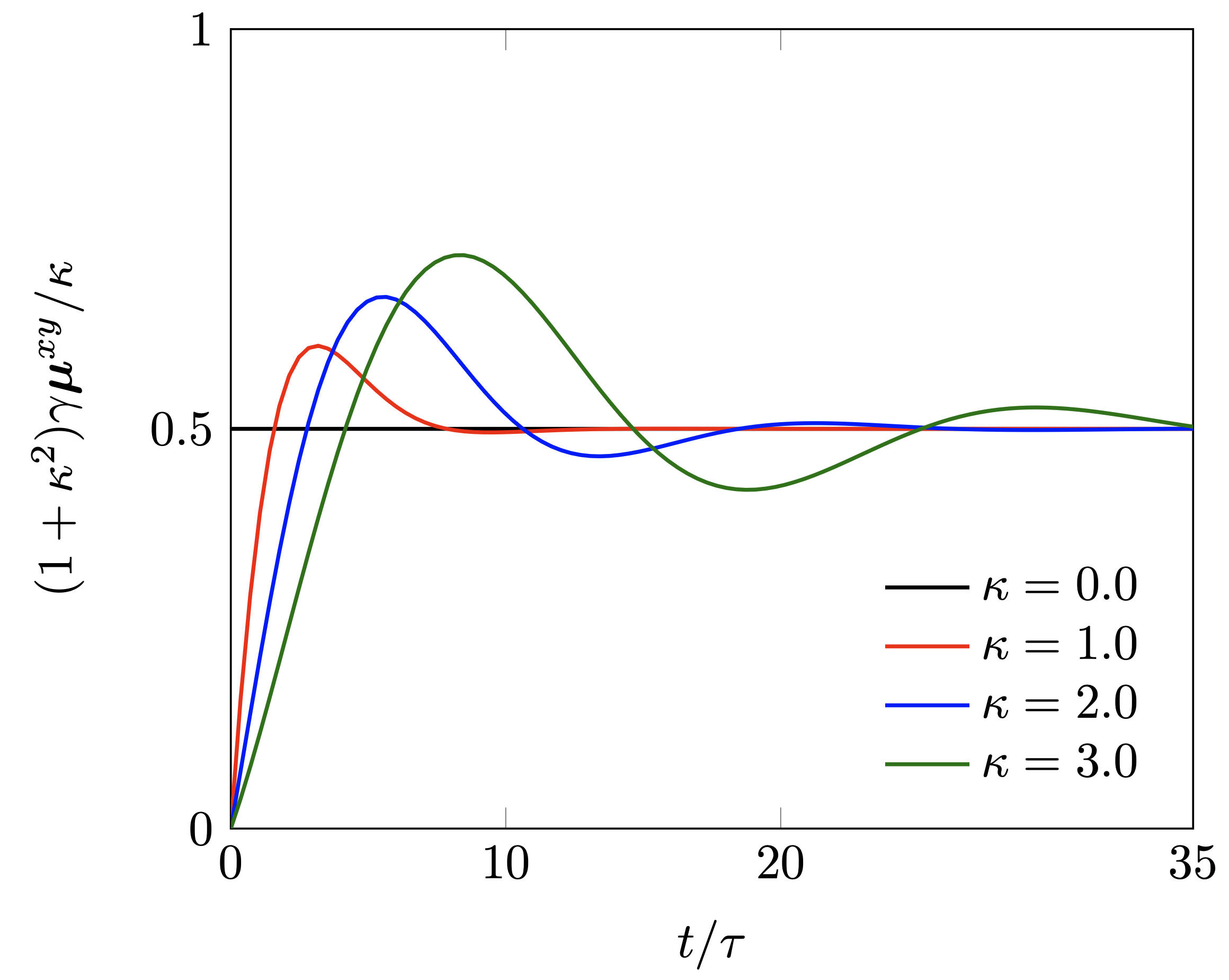}
    \caption{}
    \label{fig:second}
\end{subfigure}    
\caption{(a) Plot of the diagonal component of the mobility tensor, $(1+\kappa^2)\gamma\boldsymbol{\mu}_{xx(t)}$, as a function of the dimensionless time $t/\tau$ and $\kappa$. (b) Plot of the off-diagonal component of the mobility tensor, $(1+\kappa^2)\beta\boldsymbol{\mu}_{xy(t)}/\kappa$, as a function of the dimensionless time $t/\tau$ and $\kappa$.}
\label{fig_mobility}
\end{figure}

The mobility tensor follows analogously from the time integral of the VACT:
\begin{equation}\label{D_DIMER}
\begin{split}
    (1+\kappa^2)\gamma\mu_{xx}(t) &= 1 - \frac{1}{2}
    \left[1 - e^{-t/\tau_\kappa}\Bigl(\cos(t\kappa/\tau_\kappa) - \kappa\sin(t\kappa/\tau_\kappa)\Bigr)\right], \\
    (1+\kappa^2)\gamma\mu_{xy}(t) &= \kappa - \frac{1}{2}
    \left[\kappa - e^{-t/\tau_\kappa}\Bigl(\sin(t\kappa/\tau_\kappa) - \kappa\cos(t\kappa/\tau_\kappa)\Bigr)\right], \\
    \gamma\mu_{zz}(t) &= 1 - \frac{1}{2}\left(1 - e^{-t/\tau}\right).
\end{split}
\end{equation}
These components are shown in Fig.~\ref{fig_mobility}. As $\kappa$ increases, the 
oscillatory character of $\mu_{xx}$ and $\mu_{xy}$ becomes more pronounced, and 
$\mu_{xx}(t)$ can become \emph{negative} -- a manifestation of absolute negative mobility, 
directly traceable to the oscillatory FACT through Eq.~\eqref{VACT_FACT_OVER}.

\begin{figure}[h!]
    \centering
    \includegraphics[width=0.5\linewidth]{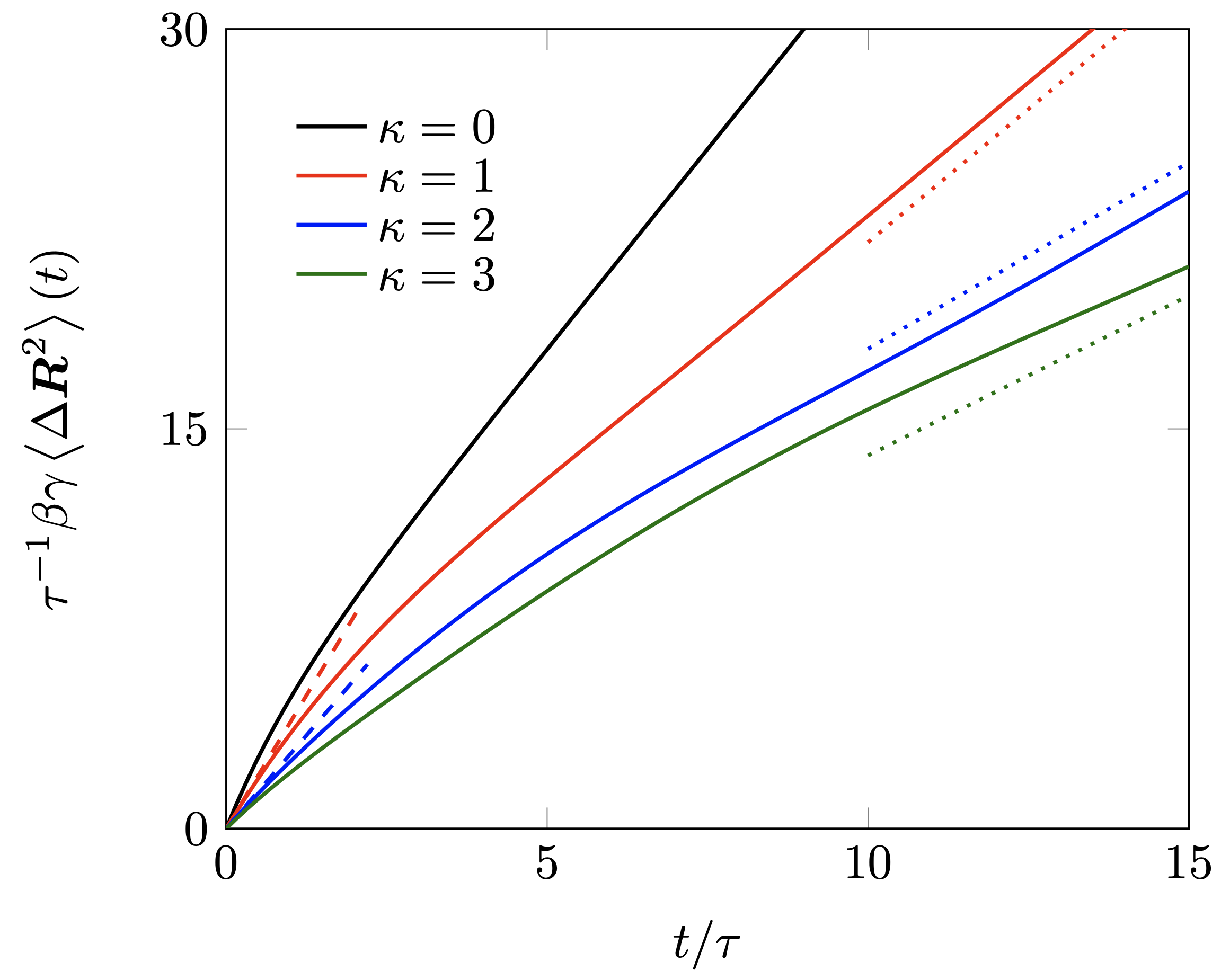}
    \caption{Plot of the mean square displacement, $\tau^{-1}\beta\gamma\left<\vert \Delta \boldsymbol R\vert^2_{(t)} \right>$, as a function of the dimensionless time $t/\tau$ for different values of $\kappa$. The dashed and the dotted lines represent respectively the short time and the long time expansion.}
    \label{MSD_same}
\end{figure}

The mean square displacement follows from Eq.~\eqref{DR}:
\begin{equation}\label{R_DIMER}
\begin{split}
    \bigl\langle|\Delta\boldsymbol{R}|^2(t)\bigr\rangle = 
    &6\frac{k_BT}{\gamma}t\left\{\frac{1+\kappa^2/3}{1+\kappa^2} 
    - \frac{1}{3}\frac{1}{1+\kappa^2}\left[1 - \frac{1+\kappa^2}{t/\tau}
    \left(1-e^{-t/\tau_\kappa}\cos(t\kappa/\tau_\kappa)\right)\right]\right\} \\
    - &\frac{k_BT}{\gamma}t\left[1 - \frac{1}{t/\tau}\left(1-e^{-t/\tau}\right)\right],
\end{split}
\end{equation}
shown in Fig.~\ref{MSD_same}. At long times, the in-plane and out-of-plane contributions 
separate cleanly: the in-plane part of the diffusion coefficient is reduced to 
$2(k_BT/\gamma)/(1+\kappa^2)$, monotonically decreasing from $2k_BT/\gamma$ at $\kappa=0$ 
toward zero as $\kappa\to\infty$, while the $z$-direction contributes the field-independent 
value $k_BT/\gamma$. The total long-time growth rate is therefore
\begin{equation}
    \bigl\langle|\Delta\boldsymbol{R}|^2(t)\bigr\rangle 
    \xrightarrow{t\to\infty} \frac{k_BT}{\gamma}\,\frac{3+\kappa^2}{1+\kappa^2}\,t,
\end{equation}
which decreases monotonically from $3(k_BT/\gamma)\,t$ at $\kappa=0$ to $(k_BT/\gamma)\,t$ 
as $\kappa\to\infty$. Unlike the opposite-charge case, here the suppression of diffusion by 
the magnetic field \emph{persists} at long times: it is not recovered, and originates 
entirely from the in-plane oscillatory coupling that survives in the relative coordinate.

Taken together, the two cases illustrate the role of relative charge sign in determining 
whether the Lorentz force leaves a lasting imprint on the dynamics of an interacting pair. 
For opposite charges, the rotational coupling cancels in the relative coordinate, and the 
magnetic field produces only a transient suppression of mobility and diffusion that is 
fully recovered at long times. For equal charges, the coupling survives, generating an 
oscillatory FACT, regions of negative mobility, and a persistent, $\kappa$-dependent 
suppression of long-time self-diffusion in the plane perpendicular to the field. We note 
that the \emph{sign} of this effect is model-dependent: here the in-plane diffusion is 
suppressed, whereas an enhancement of self-diffusion has been reported for interacting 
particles with hard-sphere repulsion~\cite{Osc_Forc_Autcorr, Coll_Enh_Self_Diff}. The general 
relations derived in Secs.~II and III accommodate both outcomes; which one is realized 
depends on the detailed structure of the interaction, consistent with 
Refs.~\cite{shinde2022strongly, Osc_Forc_Autcorr}.

\newpage
\section{Conclusion}

The Introduction posed a basic physical question: since the Lorentz force is 
velocity-dependent, and velocity is itself strongly randomized by collisions with the 
solvent, does the magnetic field leave any measurable trace in the configurational-space 
dynamics of a Brownian particle, or is its effect simply averaged away? In this work we 
have answered this question constructively, by deriving a general relation between the 
velocity and force autocorrelation tensors (VACT and FACT) in the presence of a magnetic 
field, valid for arbitrary conservative interactions which obey radial symmetry. The derivation rests entirely on the 
time-symmetry structure of the underlying equations of motion, first for the full Langevin 
dynamics and subsequently in the overdamped (Brownian) limit.

The central qualitative novelty of the overdamped relation is the following: in the 
presence of a magnetic field, the diagonal part of the VACT -- the physically measurable 
quantity that controls the mean square displacement -- couples not only to the diagonal 
part of the FACT, as in the field-free case, but also to its \emph{off-diagonal}, 
antisymmetric part. This off-diagonal component encodes the chirality of force 
correlations induced by the field, and is dynamically invisible to self-diffusion when 
$B=0$. Once time-reversal symmetry is broken, however, it becomes directly relevant. The 
degree to which this mixing occurs is governed by the single dimensionless parameter 
$\kappa = \omega_c\tau_{th}$, the product of the cyclotron frequency $\omega_c$ and the 
thermalization time $\tau_{th}$: $\kappa$ controls both the magnitude of the coupling and, 
as we showed explicitly, the relative weighting of diagonal and off-diagonal FACT 
contributions to the VACT.

We validated and illustrated this formalism on an exactly solvable model: a dimer of 
charged particles bound by a harmonic potential. The two cases of relative charge 
considered reveal sharply different physics. For oppositely charged particles, the 
rotational coupling induced by the field cancels in the relative coordinate, and the 
magnetic field produces only a transient effect: the mobility and diffusion are suppressed 
at short times but fully recover, becoming field-independent, at long times -- reproducing 
and extending the findings of Ref.~\cite{shinde2022strongly}. For particles of equal 
charge, by contrast, the rotational coupling survives in the relative coordinate, the FACT 
acquires a genuine oscillatory behavior in time, the mobility tensor develops regions of 
absolute negative mobility, and the in-plane self-diffusion is suppressed \emph{persistently} 
at long times. This second case is the minimal, exactly solvable realization of the general 
mechanism established in Sec.~III: an oscillatory off-diagonal FACT feeding directly into 
the long-time diffusive behavior of the particle. Such oscillatory force autocorrelations, 
and their consequences for mobility and diffusion, have also been reported in related 
contexts~\cite{Osc_Forc_Autcorr, Mobility, Reversal}, and the present work provides a 
general, model-independent framework within which these observations can be understood.

A natural extension of this work is to interacting many-body systems beyond the solvable 
dimer, where oscillatory force autocorrelations have already been observed numerically for 
particles with hard-sphere repulsion~\cite{Osc_Forc_Autcorr}. The general relation derived here 
provides the appropriate framework to interpret such results and to predict the sign and 
magnitude of the resulting modification of self-diffusion from the structure of the FACT 
alone, without requiring a model-specific calculation. The application of this formalism to 
dusty plasmas, where strongly charged, magnetized particles are routinely realized 
experimentally, remains an open and promising direction. Quantitative comparison with such 
systems will require going beyond the present treatment, which neglects the dynamical 
response of the background fluid; incorporating these effects into the present symmetry-based 
framework is left for future work.

\section{Appendix A: Detailed balance for the Langevin dynamics}

In this appendix we justify, at the level of the Fokker-Planck equation, why the 
symmetry relations derived in Sec.~I for Hamiltonian dynamics extend to the stochastic 
Langevin dynamics of Eq.~\eqref{langevin_B}.

\subsection{Fokker-Planck equation}

The probability density $P(\boldsymbol{X},t)$ associated with Eq.~\eqref{langevin_B} obeys 
the Fokker-Planck equation
\begin{equation}\label{FP}
    \partial_t P = \LFP{B}\,P, 
    \qquad \LFP{B} = \Liou{B} + \Lir,
\end{equation}
where $\Liou{B}$ is the reversible (streaming) part already introduced in Sec.~I, and the 
irreversible part is
\begin{equation}
    \Lir[\,\cdot\,] = \frac{\gamma}{m}\nabla_{\boldsymbol{v}}\cdot
    \Bigl(\boldsymbol{v}\,[\,\cdot\,] + \frac{k_BT}{m}\nabla_{\boldsymbol{v}}[\,\cdot\,]\Bigr).
\end{equation}
The steady-state solution of Eq.~\eqref{FP} is the Boltzmann distribution $P_{eq}(\boldsymbol{X})$.

\subsection{Detailed balance under time reversal}

In the deterministic case, the symmetry relations~\eqref{TRS_1} and~\eqref{TRS_2} followed 
from the invariance of the phase-space trajectories under $\Top$ and $\Rop{j}$. In the 
stochastic case, the analogous statements are detailed-balance relations for the transition 
probability density $P(\boldsymbol{X},t\,|\,\boldsymbol{X}_0)\atB$, the probability of 
reaching state $\boldsymbol{X}$ at time $t$ starting from $\boldsymbol{X}_0$, at fixed 
magnetic field $B$ along $\hat{\mathbf{z}}$. Under the time-reversal operator $\Top$, the 
relevant statement is
\begin{equation}\label{detailed_balance_1}
    P(\boldsymbol{X},t\,|\,\boldsymbol{X}_0)\atB\,P_{eq}(\boldsymbol{X}_0) 
    = P(\Top\boldsymbol{X}_0,t\,|\,\Top\boldsymbol{X})\atmB\,P_{eq}(\Top\boldsymbol{X}).
\end{equation}
This relation holds if and only if the generator satisfies
\begin{equation}\label{detailed_balance_1_cond}
    \LFP{B}\,P_{eq}(\boldsymbol{X})\,[\,\cdot\,] 
    = P_{eq}(\Top\boldsymbol{X})\,
    \bigl(\hat{L}_{(\Top\boldsymbol{X},-B)} + \hat{L}^{ir}_{(\Top\boldsymbol{X})}\bigr)^\dagger[\,\cdot\,],
\end{equation}
where the dagger denotes the adjoint operator. This condition can be verified by direct 
inspection.

\subsection{Detailed balance under the spatial reflections}

An analogous statement holds for each of the spatial reflection operators $\Rop{j}$, 
$j=1,\ldots,4$, introduced in Sec.~I, now at fixed magnetic field:
\begin{equation}\label{detailed_balance_2}
    P(\boldsymbol{X},t\,|\,\boldsymbol{X}_0)\atB\,P_{eq}(\boldsymbol{X}_0) 
    = P(\Rop{j}\boldsymbol{X}_0,t\,|\,\Rop{j}\boldsymbol{X})\atB\,P_{eq}(\Rop{j}\boldsymbol{X}),
    \qquad j=1,\ldots,4,
\end{equation}
valid provided
\begin{equation}\label{detailed_balance_2_cond}
    \LFP{B}\,P_{eq}(\boldsymbol{X})\,[\,\cdot\,] 
    = P_{eq}(\Rop{j}\boldsymbol{X})\,
    \bigl(\hat{L}_{(\Rop{j}\boldsymbol{X},B)} + \hat{L}^{ir}_{(\Rop{j}\boldsymbol{X})}\bigr)^\dagger[\,\cdot\,],
\end{equation}
again verified by direct inspection.

\subsection{Consequences for correlation functions}

Applying Eqs.~\eqref{detailed_balance_1} and~\eqref{detailed_balance_2} to the evaluation 
of equilibrium time correlation functions reproduces, respectively, the symmetry relations 
of Eqs.~\eqref{TRS_1} and~\eqref{TRS_2}. Consequently, the tensor structure of the VACT and 
FACT derived in Sec.~I from these symmetries -- and used throughout the rest of this work 
-- holds equally for the Langevin dynamics governed by Eq.~\eqref{langevin_B}. A detailed 
discussion of detailed balance for Langevin dynamics can be found in Ref.~\cite{FP}, 
Chapter VI.

\bibliography{Bibliography}% Produces the bibliography via BibTeX.

\end{document}